# Application of Artificial Neural Networks in Predicting Abrasion Resistance of Solution Polymerized Styrene-Butadiene Rubber Based Composites

Hao Li
1. College of Chemistry, Sichuan University,
Chengdu, Sichuan 610064, China
E-mail: lihao_chem_92@hotmail.com

Dazuo Yang[2,3,*], Fudi Chen[3], Yibing Zhou[3], Zhilong Xiu[3]
2. Key Laboratory of Marine Bio-resources Restoration and Habitat Reparation in Liaoning Province, Dalian Ocean University
Dalian 116023, P. R. China
3. College of Life science and Technology, Dalian University of Technology
Dalian 116023, P. R. China
* E-mail: dzyang1979@hotmail.com

*Abstract*-Abrasion resistance of solution polymerized styrene-butadiene rubber (SSBR) based composites is a typical and crucial property in practical applications. Previous studies show that the abrasion resistance can be calculated by the multiple linear regression model. In our study, considering this relationship can also be described into the non-linear conditions, a Multilayer Feed-forward Neural Networks model with 3 nodes (MLFN-3) was successfully established to describe the relationship between the abrasion resistance and other properties, using 23 groups of data, with the RMS error 0.07. Our studies have proved that Artificial Neural Networks (ANN) model can be used to predict the SSBR-based composites, which is an accurate and robust process.

*Keywords*-solution polymerized styrene-butadienerubber; abrasion resistance; artificial neural networks; multilayer feed-forward neural networks; prediction.

## I. INTRODUCTION

Rubber, a kind of elastic materials, is now be widely used in the practical applications and researches [1-2]. Auto tyre is a typical rubber matrix composite structures, which is a crucial property of the automobile, sustaining the weight of the vehicles for shock absorption [3]. Abrasion resistance is one of the most significant performances of these rubber materials, which is highly correlate with the working life of rubber items [4]. Therefore, it is of great importance that to study the abrasion resistance of rubber materials. In practical applications and researches, however, because of the complexity of the norms and operations, the precise abrasion resistance of rubber materials are difficult to determine. In previous studies, B. Wang and his co-workers [5] developed a multiple linear regression model to estimate the abrasion resistance of solution polymerized styrene-butadiene rubber (SSBR), with precise results. In our research work, we aimed at studying the non-linear relationships of abrasion resistance of SSBR and other properties, for establishing a precise and robust Artificial Neural Networks (ANN) model to predict the precise values of SSBR's abrasion resistance.

## II. ARTIFICIAL NEURAL NETWORKS

### A. Fundamental of ANN models

Network model is widely used in many areas, such as signal and image processing [6-7], wireless sensor network [8-10], biological modeling [11-13]. A neural network is composed of an interconnected group of artificial neurons, and a connectionist way is taken to process information for it. In most circumstances, an artificial neural network (ANN) is an adaptive system that is equipped to be adapting continuously to new data and learning from the accumulated experience [14] and noisy data. Apart from that, the system structure can be changed based on external or internal information that flows through the network during the learning phase. Meanwhile, essential information can be abstract from data or model complex relationships between inputs and outputs.

From Figure 1, the main structure of the artificial neural network (ANN) is made up of the input layer and the output layer. The input variables are introduced to the network by the input layer [15]. Also, the response variables with predictions, which stand for the output of the nodes in this certain layer, provided by the network .Additionally, the hidden layer is included. The type and the complexity of the process or experimentation usually iteratively determine the optimal number of the neurons in the hidden layers [16].

### B. Training process of ANN models

The data of abrasion and other mechanical properties of SSBR-based rubber composites are obtained from B. Wang's work [5], which is shown on Table 1.

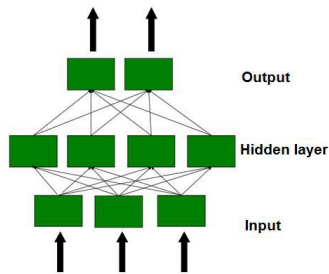

Figure 1. A schematic view of artificial neural network structure.

TABLE I. ABRASION AND OTHER MECHANICAL PROPERTIES OF SSBR-BASED RUBBER COMPOSITES

| Sample | Akron abrasion (1.61km) /cm³ | Shore A hardness | Modulus at 100% / MPa | Modulus at 300% / MPa | Modulus at 300% /modulus at 100% | Tensile strength / MPa | Elongation at break /% | Tear strength / (kN · m⁻¹) | Permanent set /% |
|---|---|---|---|---|---|---|---|---|---|
| 1 | 0.18 | 62 | 2.1 | 11.5 | 5.5 | 18.1 | 417 | 38.3 | 12 |
| 2 | 0.21 | 62 | 2.1 | 10.7 | 5.1 | 17.5 | 427 | 41.8 | 12 |
| 3 | 0.22 | 70 | 2.0 | 10.8 | 5.4 | 15.5 | 456 | 47.2 | 12 |
| 4 | 0.29 | 64 | 2.6 | 11.2 | 4.3 | 16.7 | 405 | 40.3 | 12 |
| 5 | 0.31 | 64 | 2.4 | 10.1 | 4.2 | 15.7 | 414 | 41.9 | 12 |
| 6 | 0.13 | 67 | 1.5 | 7.6 | 5.1 | 24.3 | 698 | 53.6 | 20 |
| 7 | 0.13 | 63 | 1.4 | 8.3 | 5.9 | 25.3 | 680 | 57.0 | 20 |
| 8 | 0.09 | 70 | 1.7 | 8.9 | 5.2 | 22.0 | 628 | 42.3 | 20 |
| 9 | 0.08 | 63 | 1.5 | 9.3 | 6.2 | 20.4 | 568 | 42.0 | 16 |
| 10 | 0.10 | 71 | 2.5 | 12.3 | 4.9 | 21.5 | 475 | 40.6 | 20 |
| 11 | 0.17 | 69 | 2.3 | 10.6 | 4.6 | 17.2 | 470 | 46.5 | 12 |
| 12 | 0.07 | 65 | 2.4 | 11.7 | 4.9 | 27.6 | 502 | 53.0 | 16 |
| 13 | 0.16 | 70 | 2.5 | 12.1 | 4.8 | 20.1 | 421 | 49.1 | 12 |
| 14 | 0.34 | 60 | 1.4 | 5.9 | 4.2 | 16.6 | 557 | 41.7 | 16 |
| 15 | 0.18 | 60 | 1.8 | 12.0 | 6.7 | 18.8 | 405 | 36.5 | 8 |
| 16 | 0.20 | 60 | 1.9 | 12.7 | 6.7 | 20.1 | 357 | 34.0 | 12 |
| 17 | 0.23 | 60 | 1.9 | 12.9 | 6.8 | 19.4 | 397 | 34.3 | 16 |
| 18 | 0.26 | 62 | 2.1 | 12.6 | 6.0 | 16.0 | 336 | 33.3 | 12 |
| 19 | 0.35 | 62 | 1.5 | 9.2 | 6.1 | 16.9 | 475 | 32.2 | 20 |
| 20 | 0.29 | 67 | 1.9 | 10.1 | 5.3 | 19.6 | 328 | 44.8 | 16 |
| 21 | 0.34 | 64 | 1.8 | 6.7 | 4.5 | 19.9 | 577 | 35.0 | 10 |
| 22 | 0.19 | 61 | 2.4 | 12.1 | 5.0 | 18.7 | 427 | 45.1 | 16 |
| 23 | 0.21 | 63 | 2.7 | 12.5 | 4.6 | 19.2 | 436 | 49.0 | 16 |

The ANN models were developed by two kinds of typical models : General Regression Neural Networks (GRNN) [17] and Multilayer Feed-forward Neural Networks (MLFN) [18]. The nodes of MLFN models are set from 2 to 16, so that, we can observe the change of the RMS error. The results are the average of a series of repeated experiments, in order to ensure the robustness of the models. The training results of different ANN models are shown on Table 2:

TABLE II. RESULTS OF DIFFERENT ANN MODELS IN PREDICTING SSBR'S AKRON ABRASION

| ANN model | Trained samples | Tested samples | RMS error | Running time |
|---|---|---|---|---|
| GRNN | 18 | 5 | 0.14 | 0:00:00 |
| MLFN 2 Nodes | 18 | 5 | 0.08 | 0:00:36 |
| MLFN 3 Nodes | 18 | 5 | 0.07 | 0:00:39 |
| MLFN 4 Nodes | 18 | 5 | 0.09 | 0:00:43 |
| MLFN 5 Nodes | 18 | 5 | 0.12 | 0:00:53 |
| MLFN 6 Nodes | 18 | 5 | 0.09 | 0:01:09 |
| MLFN 7 Nodes | 18 | 5 | 0.09 | 0:01:17 |
| MLFN 8 Nodes | 18 | 5 | 0.17 | 0:01:43 |
| MLFN 9 Nodes | 18 | 5 | 0.15 | 0:02:05 |
| MLFN 10 Nodes | 18 | 5 | 0.12 | 0:02:27 |
| MLFN 11 Nodes | 18 | 5 | 0.44 | 0:02:49 |
| MLFN 12 Nodes | 18 | 5 | 0.45 | 0:04:00 |
| MLFN 13 Nodes | 18 | 5 | 0.24 | 0:05:22 |
| MLFN 14 Nodes | 18 | 5 | 0.14 | 0:07:57 |
| MLFN 15 Nodes | 18 | 5 | 0.08 | 0:18:40 |
| MLFN 16 Nodes | 18 | 5 | 0.07 | 8:15:05 |

Table 2 presents the results of different ANN models. The MLFN model with 3 nodes (MLFN-3) and MLFN model with 16 nodes (MLFN-16) have the lowest RMS error. Meanwhile, the training time will rapidly increase by increasing the nodes' number. The training time of MLFN-3 model is much longer than MLFN-16. Therefore, the MLFN model with 3 nodes is selected to predict the akron abrasion of SSBR-based rubber composites with the lowest time consuming and RMS error.

C. *Results and discussion*

Figure 2 to 4 depict the results of training process. We found that the values are concentrated and corresponded with the normal training process of MLFN-3 model, showing that the training process is correct and precise.

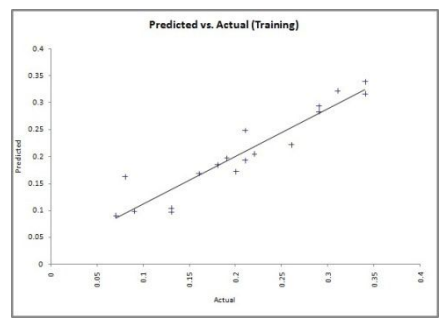

Figure 2. Comparison between predicted values and actual values during training process.

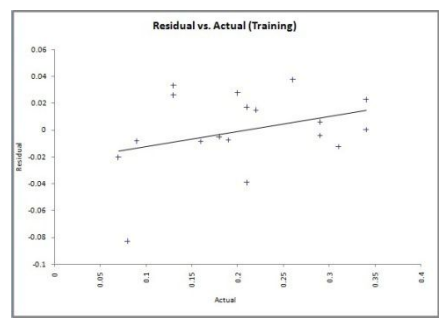

Figure 3. Comparison between residual values and actual values during training process.

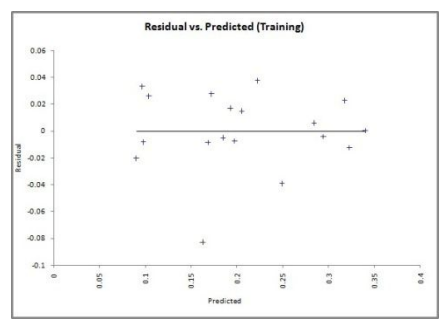

Figure 4. Comparison between residual values and predicted values during training process.

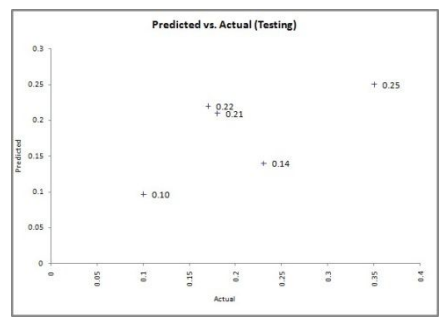

Figure 5. Comparison between predicted values and actual values during testing process.

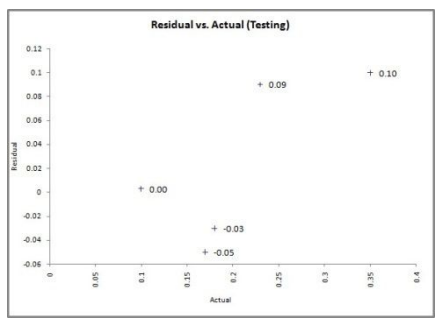

Figure 6. Comparison between residual values and actual values during testing process.

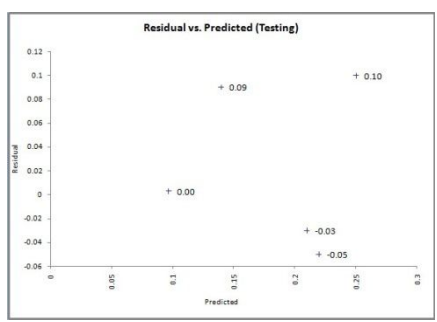

Figure 7. Comparison between residual values and predicted values during testing process.

Figure 5 to 7 depict the testing process of the MLFN-3 model. All the values in the three figures are the average values, so the model is accurate and robust.

According to the testing results, our model has been proved to be suitable and reasonable for predicting SSBR's akron abrasion. Previous studies [5] show that the relationship between abrasion resistance and other correlative properties are usually considered to be linear. However, our studies show that non-linear function is a better model for describing this relationship. We can obtain the akron abrasion of SSBR-based composites by the existing data base directly using ANN model.

## III. CONCLUSION

Our research has successfully established a Multilayer Feed-forward Neural Networks model with 3 nodes (MLFN-3), which can perfectly describe the relationship between the abrasion resistance and other properties. In future studies, we'll pay more attention to explore the relationship between different properties of solution polymerized styrene-butadiene rubber based composites.

## IV. ACKNOWLEDGEMENTS

This work was funded by the National Marine Public Welfare Research Project (No. 201305002, 201305043), and the Natural Science Foundation of Dalian, (No. 2012003219).

REFERENCE

[1] M. M.Berekaa, A.Linos, R.Reichelt, et al, "Effect of pretreatment of rubber material on its biodegradability by various rubber degrading bacteria," FEMS microbiology letters, vol. 184pp. 199-206, February 2010.
[2] Y.Shen, K.Chandrashekhara, W. F.Breig, et al, "Neural network based constitutive model for rubber material,". Rubber chemistry and technology, vol. 77, pp.257-277, February 2004.
[3] A.Birkholz, K. L.Belton, T. L.Guidotti, "Toxicological evaluation for the hazard assessment of tire crumb for use in public playgrounds,". Journal of the Air & Waste Management Association, vol. 53, pp. 903-907, April 2003.
[4] M. S.Kim, D. W.Kim, S.Ray Chowdhury, et al, "Melt‐compounded butadiene rubber nanocomposites with improved mechanical properties and abrasion resistance," Journal of applied polymer science, vol. 102, pp. 2062-2066, March 2006.
[5] B. Wang, J. H. Ma, Y.P. Wu, "Application of multiple linear regression in predicting abrasion resistance ofsolution polymerized styrene-butadiene rubber based composites (Chinese Article)," China Synthetic Rubber Industry, vol. 36, pp. 123-126, February 2013.
[6] J. Shen, et al, "Layer Depth Denoising and Completion for structured-Light RGB-D Cameras," in IEEE Conference on Computer Vision and Pattern Recognition, 2013, pp. 1187-1194.
[7] J. Shen, et al, "Virtual Mirror Rendering with Stationary RGB-D Cameras and Stored 3-D Background," IEEE Transaction on Image Processing, vol. 22(9), pp. 3433-3448, 2013.
[8] Q. Sun, et al, "A Multi-Agent-Based Intelligent Sensor and Actuator Network Design for Smart House and Home Automation," Journal of Sensor and Actuator Networks, vol. 2, pp.557-588, 2013.
[9] Q. Sun, et al, "Unsupervised Multi-Level Non-Negative Matrix Factorization Model: Binary Data Case," Journal of Information Security, vol. 3, pp. 245-250, 2012.
[10] H. Zhang, et al, "Gossip-Based Information Spreading in Mobile Networks," IEEE Transaction on Wireless Communications, vol. 12, pp.5918-5928, 2013.
[11] Y. Wang, et al, "A Conceptual Cellular Interaction Model of Left Ventricular Remodelling post-MI: Dynamic Network with Exit-Entry Competition Strategy," BMC Systems Biology, vol. 4, pp. S5, 2010.
[12] Y. Wang, et al, "Mathematical Modeling and Stability Analysis of Macrophage Activation in Left Ventricular Remodeling post-Myocardial Infarction," BMC Genomics, vol 13, pp.S21, 2012.
[13] N. Nguyen, et al, "Targeting Myocardial Infarction-Specific Protein Interaction Network Using Computational Analysis," In IEEE International Workshop on Genomic Signal Processing and Statistics, San Antonio, 2011, pp.198-201.
[14] H. Li, X.F. Liu, S.J. Yang, et al, "Prediction of Polarizability and Absolute Permittivity Values for Hydrocarbon Compounds Using Artificial Neural Networks," International Journal of Electrochemical Science, vol. 9, pp. 3725-3735, April 2014.
[15] P. Wang X. Ji, L. Zhu, et al, "Stratified analysis of the magnetic Barkhausen noise signal based on wavelet decomposition and back propagation neural network," Sensors and Actuators A: Physical, vol. 201, pp. 421-427, 2013.
[16] H. Aladag, A. Kayabasi, C. Gokceoglu, "Estimation of pressuremeter modulus and limit pressure of clayey soils by various artificial neural network models," Neural Computing and Applications, vol. 23, pp. 333-339, February 2013.
[17] Ö.Polat, T.Yıldırım, "FPGA implementation of a General Regression Neural Network: An embedded pattern classification system," Digital Signal Processing, vol. 20, pp. 881-886, March 2010.
[18] R.Pahlavan, M.Omid, A.Akram, "Energy input–output analysis and application of artificial neural networks for predicting greenhouse basil production," Energy, vol. 37, pp. 171-176, January 2012.